\documentclass[11pt,a4paper]{article}
\usepackage[utf8x]{inputenc}
\usepackage[T1]{fontenc}
\usepackage[normalem]{ulem}
\usepackage{xcolor}
\usepackage{mathptmx} 
\usepackage{multirow} 
\usepackage{float}

\usepackage{algorithm}
\usepackage{algpseudocode}

\definecolor{darkgreen}{RGB}{0,100,0}

\usepackage{comment}

\usepackage{mathtools}

\usepackage[pdftex]{graphicx} 
\usepackage{subcaption} 
\usepackage[pdftex,linkcolor=black,pdfborder={0 0 0}]{hyperref} 
\usepackage{calc} 
\usepackage{enumitem} 
\usepackage{color,soul}
\usepackage{amsfonts}  
\usepackage{amssymb}

\usepackage{authblk}

\usepackage[a4paper, lmargin=0.1\paperwidth, rmargin=0.1\paperwidth, tmargin=0.1\paperheight, bmargin=0.1\paperheight]{geometry} 

\usepackage[all]{nowidow} 
\usepackage[protrusion=true,expansion=true]{microtype} 

\usepackage{lipsum} 

\usepackage[
backend=biber,
style=numeric,
sorting=none
]{biblatex}
\usepackage{bm}
\hypersetup{ 	
pdfsubject = {},
pdftitle = {},
pdfauthor = {}
}

\begin{document}
\date{}

\title{Interpretable Meta-Learning for Multi-Objective Chemical Search}

\author[1,2]{Antonio Varagnolo}
\author[1]{Yulia Pimonova}
\author[3]{Michael G Taylor}
\author[2,*]{Raphaël Pestourie}
\author[1,*]{Nicholas Lubbers}

\affil[1]{Computing and Artificial Intelligence Division, Los Alamos National Laboratory, Los Alamos, NM 87545, USA}

\affil[2]{School of Computational Science and Engineering, Georgia Institute of Technology, Atlanta, GA 30332, USA}

\affil[3]{Theoretical Division, Los Alamos National Laboratory, Los Alamos, NM 87545, USA}


\maketitle
\renewcommand{\thefootnote}{}
\footnotetext{$^*$Corresponding authors: \texttt{rpestourie3@gatech.edu}, \texttt{nlubbers@lanl.gov}}
\renewcommand{\thefootnote}{\arabic{footnote}}
\begin{abstract}
    \noindent Navigating the vast space of synthetically accessible molecules demands surrogate models that are interpretable and capable of handling multiple competing objectives at the same time. Deep learning approaches struggle to satisfy them under the computational constraints of quantum-level chemistry. Here, we introduce a modular pipeline that combines interpretable linear meta-learning models and adaptive-confidence uncertainty quantification into an Efficient Global Optimization (EGO) framework for multi-objective molecular discovery. For the first time, linear meta-learning is deployed in a multi-objective chemical search setting: by training across chemical objectives and cheap auxiliary properties, the meta-learned surrogates acquire transferable chemical knowledge that adapts rapidly to new objectives from limited data. Evaluated empirically on a real large scale search for spin-crossover metal–organic complexes, the baseline performs 78\% worse in Pareto sense than the meta-learning alternative. We also address the calibration challenges inherent to active search. Since optimal candidates typically lie precisely in the distributional tails, standard uncertainty estimates fail. We introduce an adaptive confidence-tuning algorithm that dynamically recalibrates the exploration–exploitation trade-off as the molecular search evolves. Empirically, dynamic confidence tuning further dominates over 50\% of the statically calibrated front.
\end{abstract}

\section{Introduction}

Molecular discovery underpins advances in drug design, materials science, and energy technology, yet the space of synthetically accessible molecules is estimated to exceed $10^{60}$ candidates~\cite{reymond2015chemical}.
Searching this combinatorial vastness is fundamentally intractable by exhaustive enumeration: simply enumerating a meaningful fraction of candidate structures is prohibitively expensive, much less evaluating them with, e.g., high-throughput computational chemistry.
Effective search therefore demands both surrogate models that can approximate costly quantum-mechanical calculations from limited data and intelligent strategies for deciding which candidates to evaluate next~\cite{tabor2018accelerating, elton2019deep, von2020exploring}. \\

\noindent Machine learning and deep learning models trained on structure-property relationships now routinely accelerate the evaluation of molecular candidates across a wide range of chemical properties~\cite{Lubbers2017HIPNN, heid2023chemprop}, and have been successfully integrated into active search pipelines~\cite{janet2019designing, stocker2020machine, chakraborty2023utilizing, janet2020accurate}. Generative models offer a complementary approach, either as standalone tools conditioned on desired properties ~\cite{gomez2018automatic, rigoni2020conditionalconstrainedgraphvariational, vignac2023digressdiscretedenoisingdiffusion, liu2024graphdiffusiontransformersmulticonditional, anstine2023generative} or embedded directly inside optimization loops~\cite{zeng2022deep, lu2026feature, kim2026harnessing}. Yet these approaches often share a common limitation: they typically demand large, task-specific datasets. Although pre-trained foundation models can reduce this cost for known property classes \cite{wadell2025foundation, mendez2022mole}, the number of instances needed are still rarely available in the early stages of scientific discovery. Active learning has been used for dataset creation in order to refine and make more robust generative models in this field \cite{antoniuk2025active}. Deep models are also expensive to train and difficult to interpret. The expensiveness of deep learning methods is undesirable given the computational constraints already set by quantum-level chemistry. Lack of interpretability sits in tension with the human-in-the-loop interactiveness that search tasks often require. Another, overlapping challenge concerns the multi-objective nature of molecular discovery. Real design problems very rarely reduce to a single objective: candidates might typically need to simultaneously satisfy constraints on reactivity, stability, synthesizability, toxicity, and cost, for instance ~\cite{fromer2023computer, fromer2024pareto, al2023examining}. Multi-objective search produces a Pareto front of non-dominated solutions, but incorporating each new objective typically requires fitting an entirely new model from scratch. Two recent advances in the field of explainable AI (XAI) can effectively tackle these limitations. Firstly, linear graphlet-based surrogates offer a compelling alternative to deep learning: they are fast to fit, data-efficient and interpretable by design. They have shown to be competitive with deep models for structure-property prediction on relevant benchmark datasets \cite{Tynes2024LinearGraphlet, pimonova2025meta}. Secondly, meta-learning~\cite{huisman2021survey, vettoruzzo2024advances, wang2024comprehensive} addresses effectively the issues that arise with multi-objective searches: by learning across a distribution of related chemical tasks, a meta-learned model acquires transferable representations that adapt rapidly to new objectives with minimal data. Meta-learning has been applied to other areas of computational chemistry such as interatomic potentials prediction~\cite{allen2024learning}, multi-objective optimization of chemical reactions ~\cite{luo2026meta}, combustion kinetics ~\cite{tao2025meta} or drug discovery ~\cite{cai2023end}. Moreover, meta-learning has shown to be able to encode chemical knowledge in learned latent representations~\cite{chen2025harnessing, pimonova2025meta}, but has yet to be deployed in the domain of multi-objective chemical search problems. Recent efforts have developed linear meta-learning models on the chemical graphlet feature space that preserve interpretability ~\cite{pimonova2025meta}. The proposed approach shares with foundation models the idea that prior chemical knowledge can be encoded once and transferred rapidly to new tasks with little data while optimizing for computational efficiency and preserving interpretability. Rich sources of such prior knowledge exist in the chemical domain. Computational chemistry community has produced large quantum-mechanical datasets~\cite{ramakrishnan2014quantum, smith2020ani1ccx} spanning diverse molecular properties. Quantum simulations routinely produce auxiliary quantities carrying chemically meaningful information that are typically discarded at surrogate training time ~\cite{katritzky2010quantitative}. \\

\noindent Another pillar of an effective search pipeline is the interplay between uncertainty quantification and candidate selection. In a large chemical space, the algorithm must decide which molecules to simulate at each iteration. Prior research has often used an uncertainty-driven active learning approach for pure exploration of the chemical space ~\cite{smith2018less, kulichenko2023uncertainty, janet2019quantitative}. Others have framed chemical search in an Expected Global Optimization (EGO) loop ~\cite{janet2020accurate}, which provides a principled basis for balancing exploitation of promising regions against exploration of uncertain ones using probabilistic surrogates ~\cite{Jones1998EGO}. Calibrating predictive uncertainty during search is, however, notoriously difficult for several reasons. Firstly, the distributions against which we fit and test our surrogates are evolving and often different from each other. Secondly, the candidates we expect to be optimal and therefore choose to evaluate often lie precisely in the tails of the molecules distribution, in which data-driven prediction and uncertainty calibration are challenging. Thirdly, ideal candidates for the property of interest could lie completely outside the seen fitting distribution as atomic substructures absent from the training set could appear at any stage of the search. Standard calibration techniques assume in-distribution queries and tend to fail otherwise, motivating adaptive strategies that tune confidence dynamically as the search evolves. \\

\noindent In this work we introduce a modular and interpretable pipeline for multi-objective molecular discovery that addresses the abovementioned challenges. The core surrogate is a linear graphlet model trained with the meta-learning formulation~\cite{pimonova2025meta}, which leverages information across all chemical objectives of the search and from low-cost secondary properties to improve generalization under data scarcity. Candidate selection is driven by Bayesian bootstrapping uncertainty quantification, complemented by a dynamic confidence-tuning algorithm that adaptively recalibrates the exploration–exploitation trade-off throughout the search. We evaluate the pipeline in an offline setting on QM9 benchmark dataset~\cite{ramakrishnan2014quantum} and in a large-scale live search for spin-crossover (SCO) metal–organic complexes, a well-studied but challenging class of materials with interesting engineering applications ~\cite{halcrow2013spin,khusniyarov2016switch, janet2018accelerating,janet2020accurate, vennelakanti2023machine}.
We demonstrate our pipeline to converge nearly two orders of magnitude faster than random search in benchmarking on QM9~\cite{ramakrishnan2014quantum}. The meta surrogate outperforms its non-meta counterpart 91\% of the time, reducing RMSE by up to 47\% at a computational overhead of only $2\times$ that of a standard linear regression.
In the live search of SCO complexes, meta-learning workflow dominates 78\% of baseline Pareto fronts while underperforming other approaches in only 18\% of cases. Similarly, dynamic confidence tuning produces Pareto fronts that dominate in 52\% of cases against the 21\% of its static counterpart. An ablation study further reveals that meta-learning acts as a chemically-aware regularizer, increasing by roughly an order of magnitude the number of molecular subgraphs (here used interchangeably with molecular graphlets) that contribute to predictions and suppressing overfitting under data scarcity. To our knowledge, this is the first work to combine meta-learning approach with multi-objective optimization to perform molecular search.

\section{Methods}

\label{sec:meth}

\begin{figure}[H]
    \centering
    \includegraphics[width=0.9\linewidth]{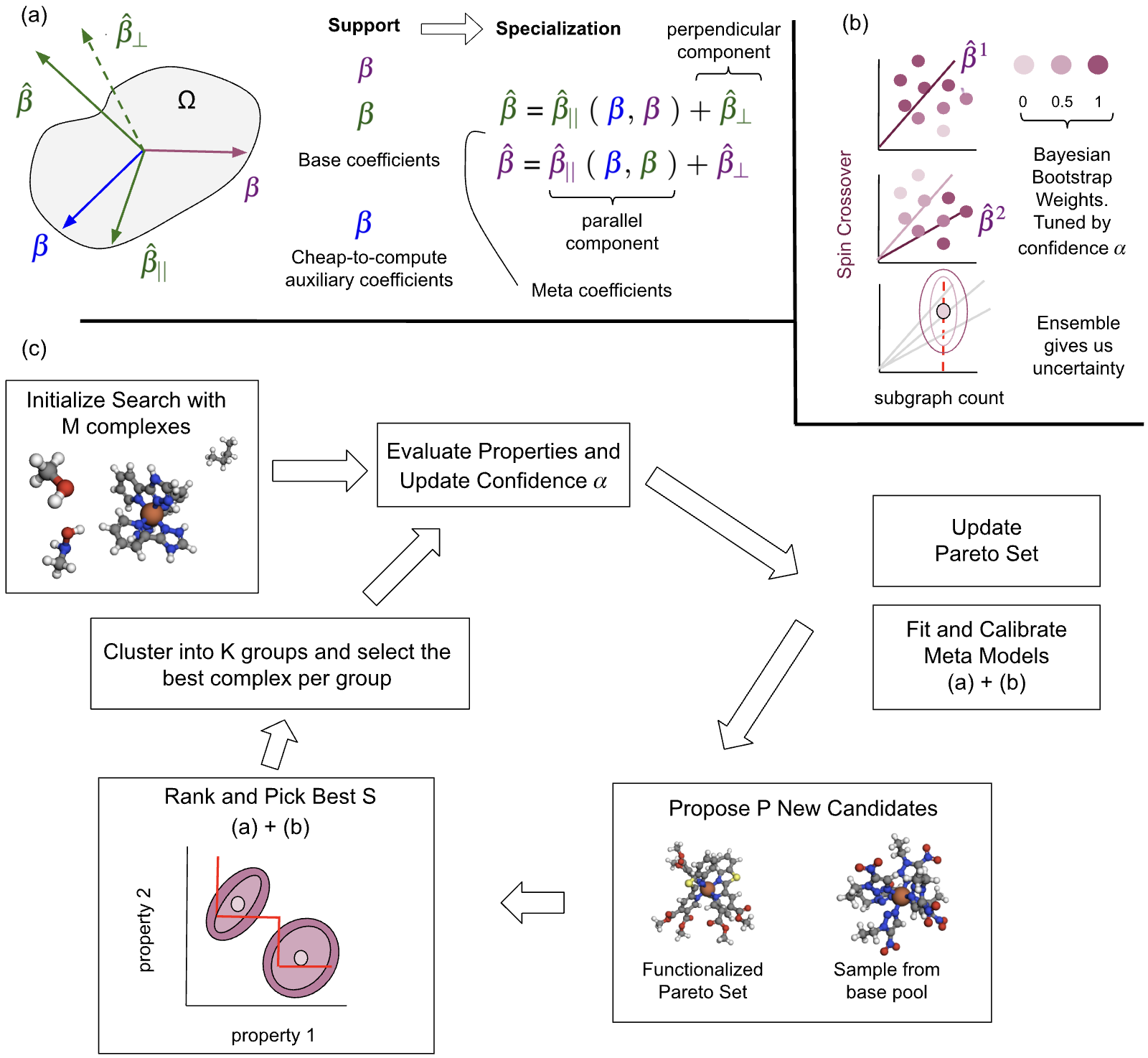}
    \caption{
    \textbf{Core components of the proposed multi-objective chemical search framework.}
    (a) The meta-learning algorithm builds new task-specific vectors as a sum of a component parallel to the subspace spanned by support-task coefficients and an orthogonal residual. 
    (b) Uncertainty is estimated with Bayesian bootstrap ensembles. The concentration parameter $\alpha$ controls the sparsity of Dirichlet-distributed weights and consequently the sparsity of the fitted parameter vector.
    (c) Iterative search loop. Starting from \(M\) evaluated complexes, the method updates the Pareto front, fits and calibrates meta-surrogates, proposes and ranks new candidates, clusters the top \(S\) candidates into \(K\) groups, evaluates representative complexes from those groups, and repeats.}
    \label{fig:pipeline}
\end{figure}

Our framework is designed as an interpretable, data-efficient search loop for multi-objective molecular discovery, as shown in Fig. ~\ref{fig:pipeline}c. Each iteration begins by encoding molecules with sparse graphlet fingerprints,~\cite{Tynes2024LinearGraphlet} against which we fit linear regressions to preserve chemical interpretability (Section~\ref{sec:linear-meta}).  Since molecular discovery often begins with little labeled data, we strengthen these task-specific linear models through a shared low-dimensional structure learned across the target objectives and cheap-to-compute auxiliary properties (Section~\ref{sec:linear-meta}). We then attach uncertainty estimates to the resulting predictions using Bayesian bootstrap ensembles with adaptive confidence parameters, providing calibrated, non-Gaussian uncertainty estimates (Section~\ref{sec:bayesian-bootstrap}). These uncertainty-aware surrogates drive an Efficient Global Optimization loop, where Pareto-aware acquisition functions select molecules that trade off predicted performance against multiple objectives and model confidence (Section~\ref{sec:ego}). The acquisition behavior is refined through a dynamic confidence strategy that adjusts the balance between exploiting promising regions and probing uncertain ones (Section ~\ref{sec:dynamic-confidence}). Exploitation is encouraged by functionalizing current Pareto-optimal complexes, whereas exploration is maintained by clustering candidates and retaining the strongest representative from each K-means cluster. Together, these components form a search pipeline that balances interpretability, data efficiency, and exploration into a single iterative strategy.

\subsection{Linear Meta-Learning for Property Prediction}

\label{sec:linear-meta}

We adopt a sparse, graphlet-based fingerprinting strategy that constructs an interpretable feature space by enumerating subgraph occurrences within the molecular graph. Each distinct graphlet defines a feature, producing a high-dimensional ($\sim 10$k features) 
but sparse descriptor vector for every molecule (Fig. ~\ref{fig:functionalization-featurization}c). The rich expressivity of the graphlet feature space allows linear models built on them to achieve predictive performance comparable to deep models while fitting orders of magnitude faster and preserving full interpretability. For a more detailed description of the linear graphlet representation and its predictive performance, we refer the reader to Ref.~\cite{Tynes2024LinearGraphlet}. \\
\\
Starting from this feature space, we adopt the linear meta-learning framework \emph{LAMeL}~\cite{pimonova2025meta} (Fig. ~\ref{fig:pipeline}a). We first clarify the terminology used throughout this section and the rest of the article. 
A \emph{support task} denotes any learning task used during meta-training, i.e. training of the meta-learning algorithm or that supports the construction a shared meta-learning space. In our case, support tasks include (i) search objectives, that form the surrogates used for ranking candidates and for which we ultimately care about performance, and (ii) auxiliary tasks corresponding to easy-to-estimate properties that are not directly optimized but help define a shared subspace structure. The total number of support tasks is denoted by $N = N_o + N_a$, where $N_o$ are search objectives and $N_a$ are auxiliary properties. A \emph{specialization task} is a typically data-scarce learning problem that is fitted adapting the meta-model built from the support tasks. In our case specialization learning tasks correspond to search objectives. Similarly, when referring to fitted linear models, we use:
\begin{itemize}
    \item \emph{base coefficients} or \emph{models} $\beta_i \in \mathbb{R}^D$ for support tasks that correspond to search objectives,
    \item \emph{auxiliary coefficients} for support tasks corresponding to auxiliary properties,
    \item \emph{support coefficients} for all support tasks,
    \item \emph{meta coefficients} or \emph{models} $\hat{\beta} \in \mathbb{R}^D$ for the final adapted model of a specialization task.
\end{itemize}
For each support task $i = 1,\dots,N$, we fit ridge regressions in the graphlet fingerprint space, obtaining vectors $\{\beta_i\}_{i=1}^N$. The central assumption of the method is that the support coefficient vectors $\{\beta_i\}_{i=1}^N$ lie in a low-dimensional subspace of $\mathbb{R}^D$ that encodes prior chemical knowledge, and can act as an inductive bias for specialization tasks. The meta coefficients $\hat{\beta}$ for a specialization task $(X^\star, y^\star)$ is constructed as
\begin{equation}
    \hat{\beta} = \beta^{\parallel} + \beta^{\perp},
\end{equation}
where the support coefficients are first centered around their mean $\bar{\beta}$ 
, the parallel component $\beta^{\parallel} = \bar{\beta} + + \sum_i c^i (\beta^i - \bar{\beta})$ 
is obtained by solving a small ridge problem of size $N \ll D$ in the subspace spanned by $\{\beta_i\}_{i=1}^N$, enforcing alignment with the low-dimensional structure shared by support tasks. The perpendicular component $\beta^{\perp}$ then fits the residual $y^\star - X^\star \beta^{\parallel}$ directly in the full fingerprint space, allowing task-specific corrections beyond the support subspace. For a more in-depth explanation of the algorithm, we refer the reader to the original linear meta-learning work~\cite{pimonova2025meta}. 
The total cost is approximately twice the cost of a single ridge regression fit. \\
\noindent In our test problems, we use as auxiliary support components the solubility properties of the conformer in water, acetone and hexane, marginal to compute once the conform is built. Other auxiliary properties could be the fast low-fidelity approximations computed with \texttt{RDKit}~\cite{rdkit}.

\subsection{Uncertainty Quantification with Bayesian Bootstrapping}

\label{sec:bayesian-bootstrap}

Unlike overparameterized models such as neural networks, where uncertainty is induced by stochastic training dynamics, the dominant sources of uncertainty in our linear surrogate models arise from misalignment between training and test data distributions (e.g. incomplete coverage of chemical structures in the training set), and limitations imposed by the maximum allowed subgraph size. Bayesian bootstrapping~\cite{Rubin1981BayesianBootstrap} provides a robust, model-agnostic framework to quantify the first of the above mentioned data-driven uncertainties. In contrast to classical bootstrapping, which relies on resampling with replacement, the Bayesian bootstrap assigns weights drawn from a Dirichlet distribution to training points. In practice, each ensemble member is fitted using a distinct set of Dirichlet-sampled weights, resulting in different effective training distributions and consequently different model coefficients $\{\beta^b\}_{b=1}^B$, where $B$ is the number of bootstrap samples (Fig. ~\ref{fig:pipeline}b). This procedure is applied uniformly to both base and meta models, yielding a total number of meta coefficients fitted equal to $N_o \times B$, and a total number of support coefficients equal to $N_o \times B + N_a$. The concentration parameter \(\alpha\) of the Dirichlet distribution controls the variability of the sampled weights and encodes a prior confidence about the observed dataset. Large values of \(\alpha\) produce weights close to uniform, corresponding to high confidence that the empirical data distribution approximates the true underlying one. Conversely, small \(\alpha\) values yield sparse weight vectors, causing individual models to rely heavily on subsets of the data and increasing the variance of ensemble predictions. Under mild regularity assumptions and for sufficiently large ensembles, a central limit theorem argument suggests that the ensemble mean prediction is insensitive to \(\alpha\), while the predictive variance increases monotonically as \(\alpha\) decreases. Bayesian bootstrapping provides a mechanism for propagating uncertainty in the empirical data distribution into predictive uncertainty.

\subsection{Multi-Objective Optimization with EGO}

\label{sec:ego}

Multi-objective optimization (MOO) of expensive black-box functions arises frequently in computational science and engineering, where several competing objectives must be optimized without explicit analytic models ~\cite{Keane2006StatImprovement, Forrester2009RecentSurrogate, Gutjahr2016SurveyNonScalar, HernandezLobato2016PESMO, MOO_Review_2018}. Pareto-based methods are a common strategy in this setting, as they aim to identify a diverse set of non-dominated solutions that approximate the Pareto front (PF), thereby preserving the trade-offs between objectives. Given two objective vectors $\mathbf{y}_1,\mathbf{y}_2\in\mathbb{R}^D$, $\mathbf{y}_1$ is said to Pareto-dominate $\mathbf{y}_2$ ($\mathbf{y}_1 \prec \mathbf{y}_2$) if it is no worse in all objectives and strictly better in at least one. Pareto dominance induces only a partial ordering, so multiple solutions may be simultaneously optimal. \\
\noindent Our optimization pipeline follows the Efficient Global Optimization (EGO) paradigm \cite{Jones1998EGO}, which iteratively combines surrogate models with acquisition functions to balance exploration and exploitation under a limited evaluation budget. EGO is often implemented with Gaussian process surrogates, which are poorly suited to our setting, where graphlet fingerprints are high-dimensional and sparse. Instead, we use Bayesian bootstrap ensembles (see Sec. ~\ref{sec:bayesian-bootstrap}) of linear meta models (see Sec. ~\ref{sec:linear-meta}).
\noindent The resulting ensembles are not Bayesian in the strict sense, but they still support acquisition-driven sampling through criteria such as probability of improvement (PoI) and expected hypervolume improvement (EHVI)~\cite{Keane2006StatImprovement, Forrester2009RecentSurrogate}.  PoI is computationally inexpensive and robust in high-dimensional spaces, but treats all non-dominated improvements equally, which can lead to conservative behavior and over-exploitation near the current front. EHVI offers a more informative selection criterion by weighting candidates according to the hypervolume they are expected to contribute beyond the current front. Its computational cost grows rapidly with both the number of objectives and the size of the Pareto front, with common formulations scaling as $O(n^D)$ for $n$ Pareto-front points in $D$ dimensions \cite{Keane2006StatImprovement, Forrester2009RecentSurrogate}. Closed-form, tractable EHVI expressions rely on Gaussian predictive distributions and independence assumptions that are not satisfied by our ensemble surrogates. Therefore, we use PoI as the acquisition score in our experiments. For lower-dimensional searches EHVI should be preferred.

\subsection{Model Confidence, Exploration, and Exploitation}

\label{sec:dynamic-confidence}

The exploration–exploitation trade-off arises at two stages of our pipeline: candidate \emph{sampling}, where molecular structures are proposed, and candidate \emph{selection}, where structures are ranked and chosen for evaluation. Exploration corresponds to sampling uncertain regions of chemical space to improve model accuracy, whereas exploitation favors candidates predicted to advance the current Pareto front.

During sampling, exploration is encouraged by drawing candidates from three distinct pools: (i) a base pool of non-Pareto molecules and (ii) Pareto molecules with novel functionalization (See Sec.~\ref{fig:functionalization-featurization}). Sampling probabilities are controlled by two parameters, \(\eta\) and \(\gamma\), such that the base pool is selected with probability \(\eta\), and the remaining probability is split between the two Pareto-derived pools. These probabilities are scheduled to decrease over generations, biasing early iterations toward exploration and later iterations toward exploitation.

Another part of our pipeline in which the exploration-exploitation decision emerges is the new molecules acquisition. A vector of confidence parameters \({\alpha}^{(t)}\) indirectly controls for the trade-off by modulating the sparsity of the Bayesian bootstrapping sampled weights. Larger values of \(\alpha^{(t)}\) correspond to higher trust in model predictions and therefore favor exploitation, whereas smaller values increase weight variability, emphasize predictive uncertainty, and promote exploration. Because the data distribution encountered during search is non-stationary and biased to look toward tail regions of chemical space, we proposed an algorithm to adapt \({\alpha}^{(t)}\) on the go to maintain calibrated uncertainty estimates throughout the search. Uncertainty calibration is quantified using the log-normalized squared error
\[
\phi^{(t)} = \log\!\left(\frac{(\epsilon^{(t)})^2}{(\sigma^{(t)})^2}\right),
\]
where \(\epsilon^{(t)}\) denotes the prediction error mean and \(\sigma^{(t)}\) the ensemble standard deviation. A well-calibrated surrogate corresponds to \(\phi^{(t)} \approx 0\). At each generation, the confidence vector is updated based on the historical sensitivity of calibration to changes in confidence, estimated from recent iterations as
\[
{\kappa}^{(t)} =
\frac{\mathrm{std}\!\left(\phi^{(t-\tau:t)}\right)}
     {\mathrm{std}\!\left({\alpha}^{(t-\tau:t)}\right)},
\]
where the statistics are computed over a fixed history window of \(\tau\) generations. The update rule is then given by
\[
\alpha^{(t+1)}
=
\alpha^{(t)}
-
\frac{\phi^{(t)}}{\kappa^{(t)}}.
\]
The feedback update drives \(\phi^{(t)}\) toward zero, keeping candidate selection in a balanced exploration--exploitation regime. To ensure numerical stability and statistically meaningful uncertainty estimates, we further constrain the confidence parameters through the effective sample size (ESS) induced by the Bayesian bootstrap. Given bootstrap weights \(w = (w_1,\dots,w_N)\) drawn from a Dirichlet distribution, the ESS is defined as
\[
\mathrm{ESS}(w) = \frac{1}{\sum_{i=1}^{N} w_i^2},
\]
which measures the number of equally weighted samples yielding the same estimator variance. 
For a symmetric Dirichlet distribution, \(\mathbf{w} \sim \mathrm{Dirichlet}(\alpha,\dots,\alpha)\), the proxy for the expected ESS admits the closed-form expression
\[
\mathbb{E}[\mathrm{ESS}] = \frac{N\alpha + 1}{\alpha + 1}.
\]
Imposing a minimum expected effective sample size \(Q\) yields the analytical lower bound
\[
\alpha \;\ge\; \frac{Q - 1}{N - Q}, \qquad N > Q,
\]
which we use to clip each component of \(\alpha^{(t+1)}\). The constraint prevents the bootstrap distribution from concentrating on too few training points and adapts automatically as the training set grows.


\section{Molecular Search Experiments}
\label{sec:experiments}

We evaluate our pipeline on two chemical search problems of increasing difficulty and computational cost. 
\subsection{QM9 Benchmark Dataset}
QM9 is a benchmark dataset containing approximately 134k small organic molecules composed of C, N, O, F, and H atoms, with up to nine heavy atoms per molecule~\cite{ramakrishnan2014quantum}. For each molecule, quantum-chemical properties are provided at the DFT level of theory. In this work, we use four of these properties as search objectives: atomization energy, zero-point energy, electronic gap, and heat capacity. QM9 provides a great testbed for benchmarking optimization strategies as the chemical space is relatively small and uniform.
Candidate evaluations reduce to table lookups, allowing us to perform many independent optimization runs and quantify empirical convergence toward the extrema of the four-dimensional objective space. The dataset is well matched to our modeling assumptions: linear graphlet-fingerprint surrogates are effective for small organic molecules, and the related nature of the target properties provides a suitable setting for the proposed meta-learning framework.

\subsection{Spin Crossover Metal-Organic Complexes}

\begin{figure}
    \centering
    \includegraphics[width=\linewidth]{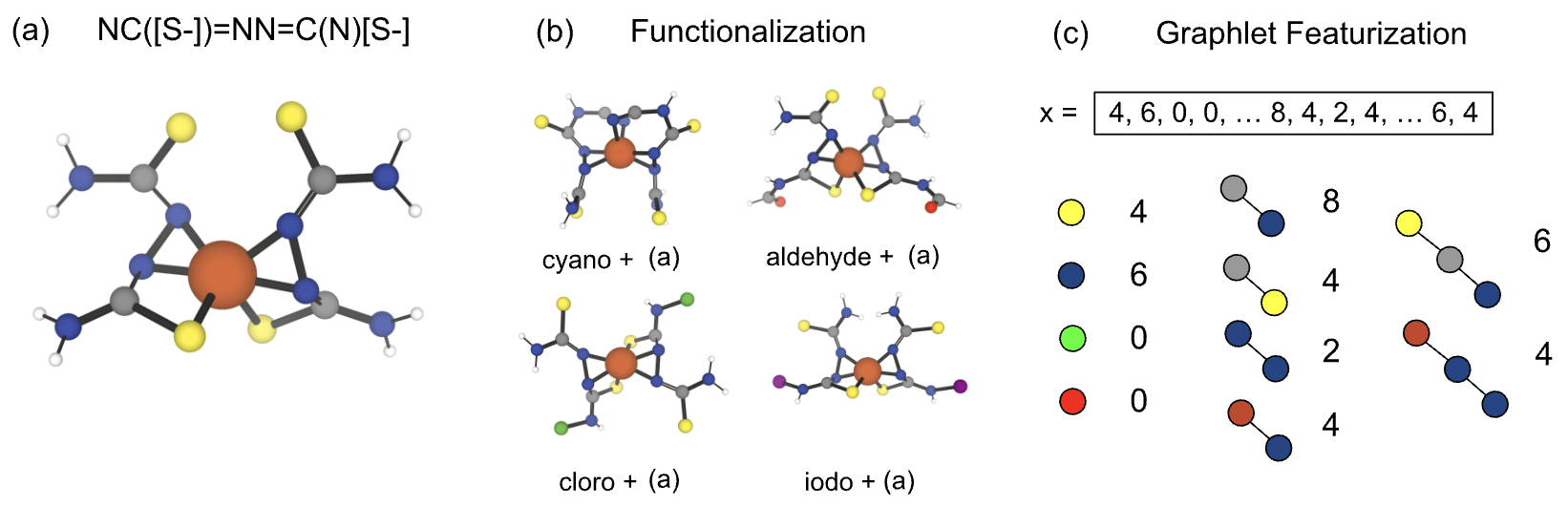}
    \caption{\textbf{Functionalization-based molecular generation and graphlet featurization.} (a) Example complex from the final Pareto front. (b) Four representative functionalizations applied to the complex in (a). 
    (c) Simplified illustration of graphleft featurization for the complex in (a). Each molecule is encoded with \texttt{minervachem} fingerprinting before being passed to the models.}
    \label{fig:functionalization-featurization}
\end{figure}

To evaluate our approach in a realistic and computationally demanding setting, we perform an online search for spin-crossover (SCO) organic–metal complexes. Spin crossover refers to the reversible transition between high-spin and low-spin electronic states of a transition-metal complex, typically induced by temperature. We focus on octahedral iron complexes with an Fe center and aim to identify candidates that both exhibit favorable spin-crossover energetics and possess good solvation-related properties. The ligand pool is constructed from organic ligands extracted from the Cambridge Structural Database (CSD) \cite{groom2016cambridge}. We apply chemically motivated filters to define a feasible ligand space. For instance, we exclude sandwich and edge-bridging ligands and retain only ligands with specific denticities and ligands that would yield total complex charge bigger than $-3$ . To further enrich the search space, we consider 28 predefined functionalizations (see Fig.~\ref{fig:functionalization-featurization}b) that can be systematically applied to the base ligands. Candidate metal–ligand complexes are then constructed in 3D space and evaluated using \texttt{Architector} ~\cite{taylor2023architector, architector_code}. \texttt{Architector} is a high-throughput Python package for modeling mononuclear coordination complexes.Starting from minimal 2D molecular graph inputs, it generates 3D conformers and estimates electronic energies for both high-spin and low-spin states using XTB \cite{bannwarth2019gfn2, bannwarth2021extended}. We define the spin-crossover energy as the difference between these two spin-state energies. In some cases, \texttt{Architector} fails to construct a valid geometry, leading to unsuccessful evaluations. As a result, the effective number of molecules added to the dataset at each iteration can be significantly smaller than the number of selected candidates. Our pipeline mitigates the risk of the pipeline steering towards unfeasible center-ligands regions by proposing functionalizations of Pareto-optimal complexes discovered in previous iterations of the search pipeline. Since these parent complexes have already been constructed successfully, their functionalized derivatives are more likely to yield new valid conformers. 

\section{Results and Discussion}
\label{sec:results}

The results from the two experiments are presented in subsections dedicated respectively to Pareto fronts results (Sec. ~\ref{sec:pareto-fronts}), surrogate models results (Sec. ~\ref{sec:models}), dynamic update of confidence results (Sec. ~\ref{sec:results-dynamic}) and ablation study of the models' coefficients (Sec. ~\ref{sec:ablation}). In section ~\ref{sec:pareto-fronts}, otherwise perfectly equal pipelines with and without meta-learning are compared. Similarly, in section ~\ref{sec:results-dynamic}, empirical pareto fronts with and without dynamic confidence updates are confronted. Section ~\ref{sec:models} focuses on the predictive performances of the surrogates, rather than the resulting pareto fronts. Finally, section ~\ref{sec:ablation} performs an ablation study on the coefficients of the meta-learning models.

\subsection{Pareto Fronts}
\label{sec:pareto-fronts}

\begin{figure}[H]
    \centering
    \includegraphics[width=1\linewidth]{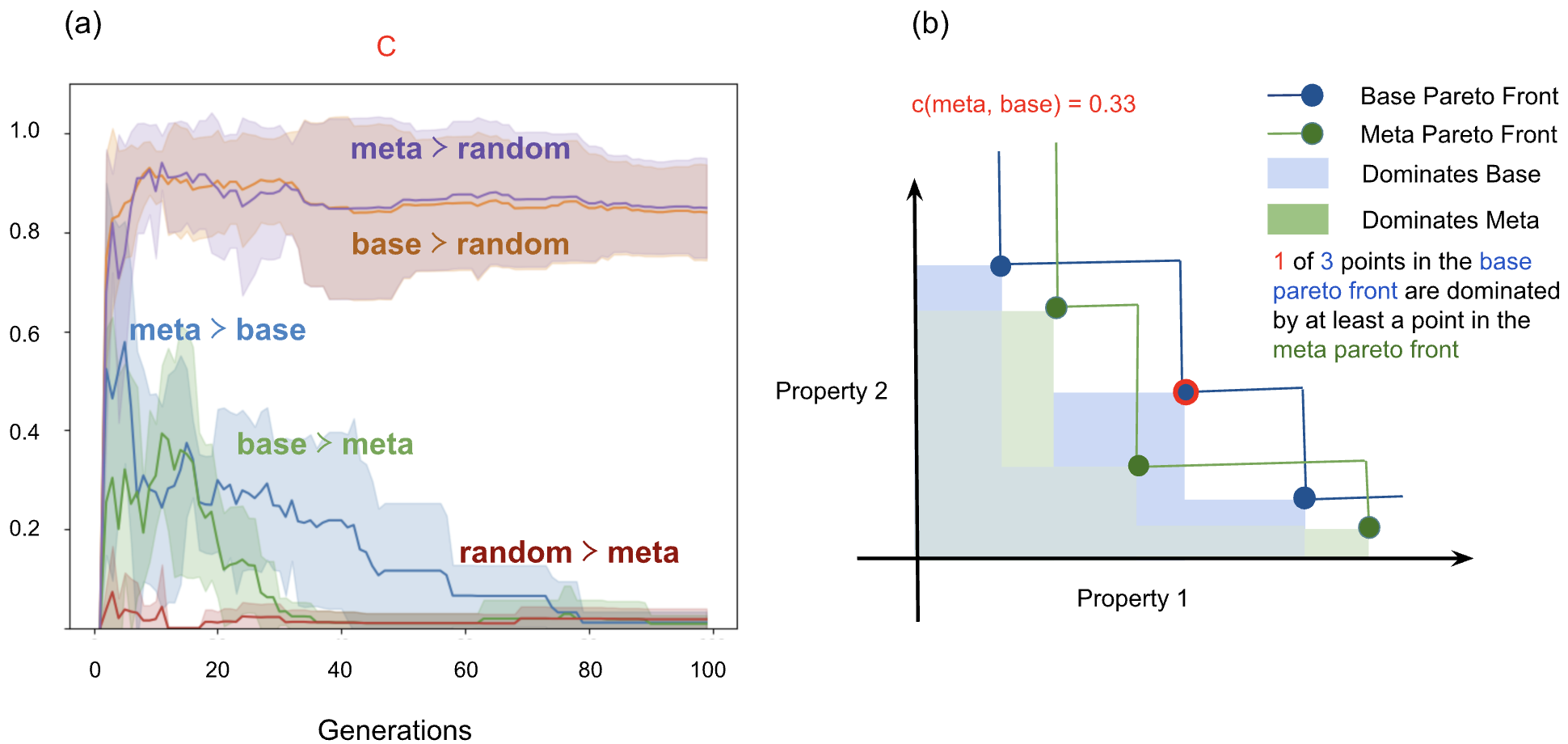}
    \caption{
    \textbf{Pareto-front coverage on QM9.}
    (a) C-metric results for the QM9 experiment, averaged over five independent runs. The random baseline selects candidates uniformly at random from the proposals generated by the search procedure, corresponding to pure exploration of the candidate space. Both the meta-learning and base models dominate the random Pareto front from the earliest stages of the search. (b) Illustration of the C-metric comparing the meta-learning Pareto front $\mathcal{P}_{\mathrm{meta}}$ against the base-model Pareto front $\mathcal{P}_{\mathrm{base}}$. The metric is defined as the fraction of points in $\mathcal{P}_{\mathrm{base}}$ that are dominated by at least one point in $\mathcal{P}_{\mathrm{meta}}$. For example, a value of 0.33 indicates that 33\% of the points in $\mathcal{P}_{\mathrm{base}}$ are dominated by at least a point in $\mathcal{P}_{\mathrm{meta}}$.}
    \label{fig:qm9_cmetric}
\end{figure}

Summarizing the results on a four dimensional Pareto front is intrinsically hard. First introduced in ~\cite{zitzler2000comparison}, the convergence, or c-metric,compares different fronts by checking in which proportion does thefront $\mathcal{P}_A$ dominate the elements in the front $\mathcal{P}_B$. We define the dominance metric of $\mathcal{P}_A$ with respect to $\mathcal{P}_B$ as:
\begin{equation}
    c(\mathcal{P}_A,\mathcal{P}_B)
    =
    \frac{1}{|\mathcal{P}_B|}
    \sum_{\mathbf{y}_B \in \mathcal{P}_B}
    \mathbb{I}
    \left(
        \exists\, \mathbf{y}_A \in \mathcal{P}_A
        \;\text{s.t.}\;
        \mathbf{y}_A \prec \mathbf{y}_B
    \right),
\end{equation}
where $\mathbf{y}_A \prec \mathbf{y}_B$ denotes Pareto dominance (i.e.\ molecule $\mathbf{x}$ is no worse than molecule $\mathbf{y}$ in all objectives and strictly better in at least one), $|\mathcal{P}_B|$ is the cardinality of $\mathcal{P}_B$, and $\mathbb{I}(\cdot)$ is the indicator function. Alternatively, $c(\mathcal{P}_A,\mathcal{P}_B)$ can be seen as the percentage of points in $\mathcal{P}_B$ that are dominated by at least one point in $\mathcal{P}_A$ (Fig. ~\ref{fig:qm9_cmetric} b). We use these metric to compare the Pareto fronts obtained by the different strategies. \\ \noindent The QM9 results are summarized in Figure~\ref{fig:qm9_cmetric}(a). The model-based Pareto fronts consistently dominate the random strategy, with C-metric values remaining above 80\% throughout the search. This can be explained by the modest size of the QM9 space and by the homogeneity of its molecules, which allow the surrogate models to become effective after only a small number of evaluations. Together, these factors lead to rapid convergence toward the true Pareto front. Comparing the two model-based pipelines, the meta-guided strategy outperforms the base (non-meta) pipeline during the early stages of the search, from generations 1 to 5, and again during mid-to-late iterations, roughly generations 20 to 80, yielding 20–30\% dominated points over the base Pareto-front. Consistent with the one-dimensional analysis, the two model-based pipelines converge to similar fronts by the end of the runs.

\begin{figure}[H]
    \centering
    \includegraphics[width=0.80\linewidth]{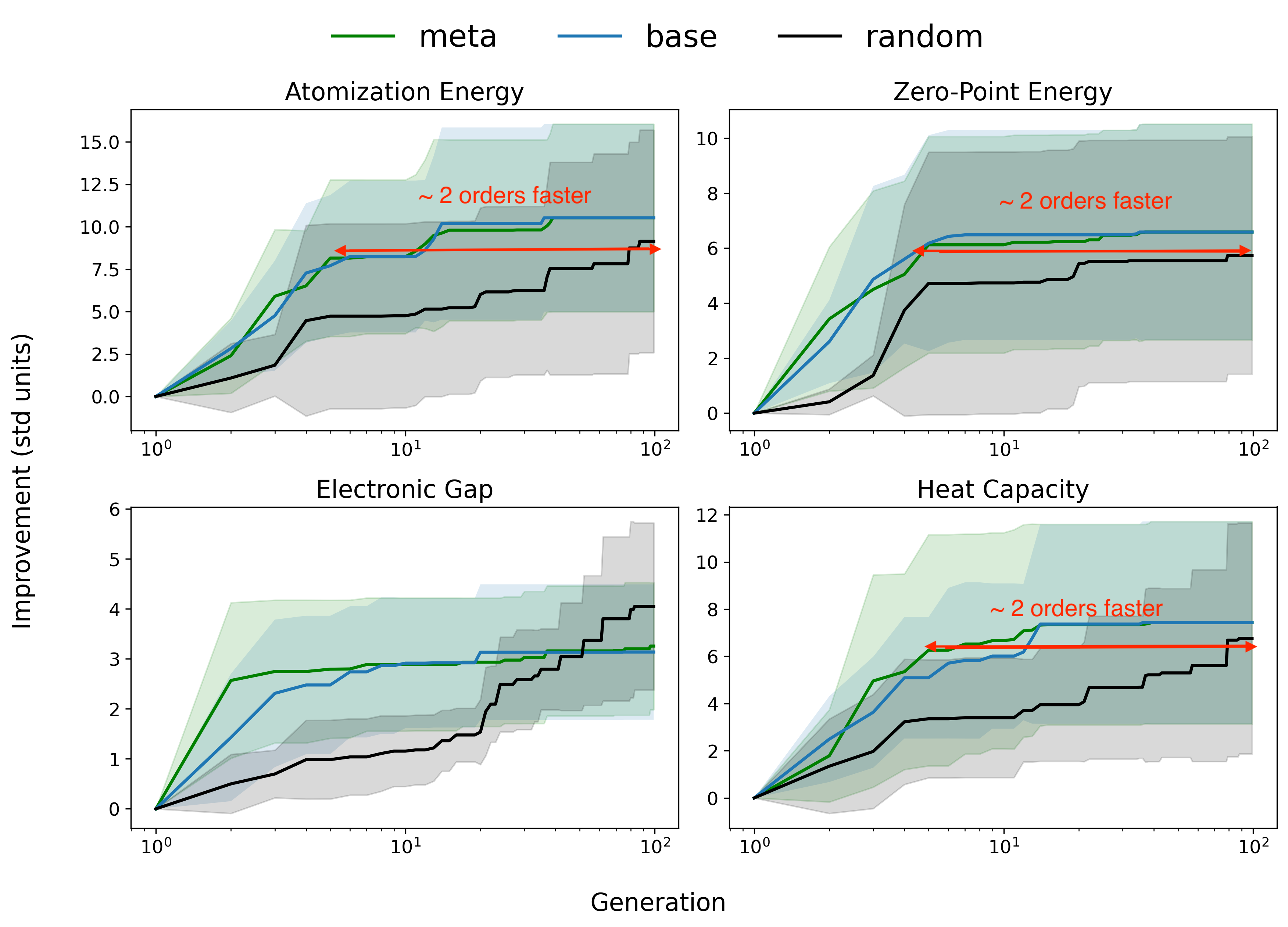}
    \caption{\textbf{Relative improvement of optima in the marginal distribution of individual QM9 properties.} Improvement at generation (t), computed as the ratio between the best objective value found by generation (t) and the best value for the same property in the initial molecular set. Curves show averages over five random seeds. 
    For the electronic gap, both model-based algorithms converge early to a suboptimal region, whereas random sampling discovers better values in the final generations.}
    \label{fig:orders-1D}
\end{figure}

In the offline QM9 experiments, our pipeline reaches target optima orders of magnitude faster than unguided exploration (Fig. ~\ref{fig:orders-1D}). For most objectives, the meta-guided search converges in roughly two orders of magnitude fewer iterations than random sampling. For three of the four target properties, both the meta-guided and base algorithms also reach better one-dimensional optima, often improving the best observed value by 1 to 1.5 units.  

\begin{figure}[H]
\centering
    \includegraphics[width=1\linewidth]{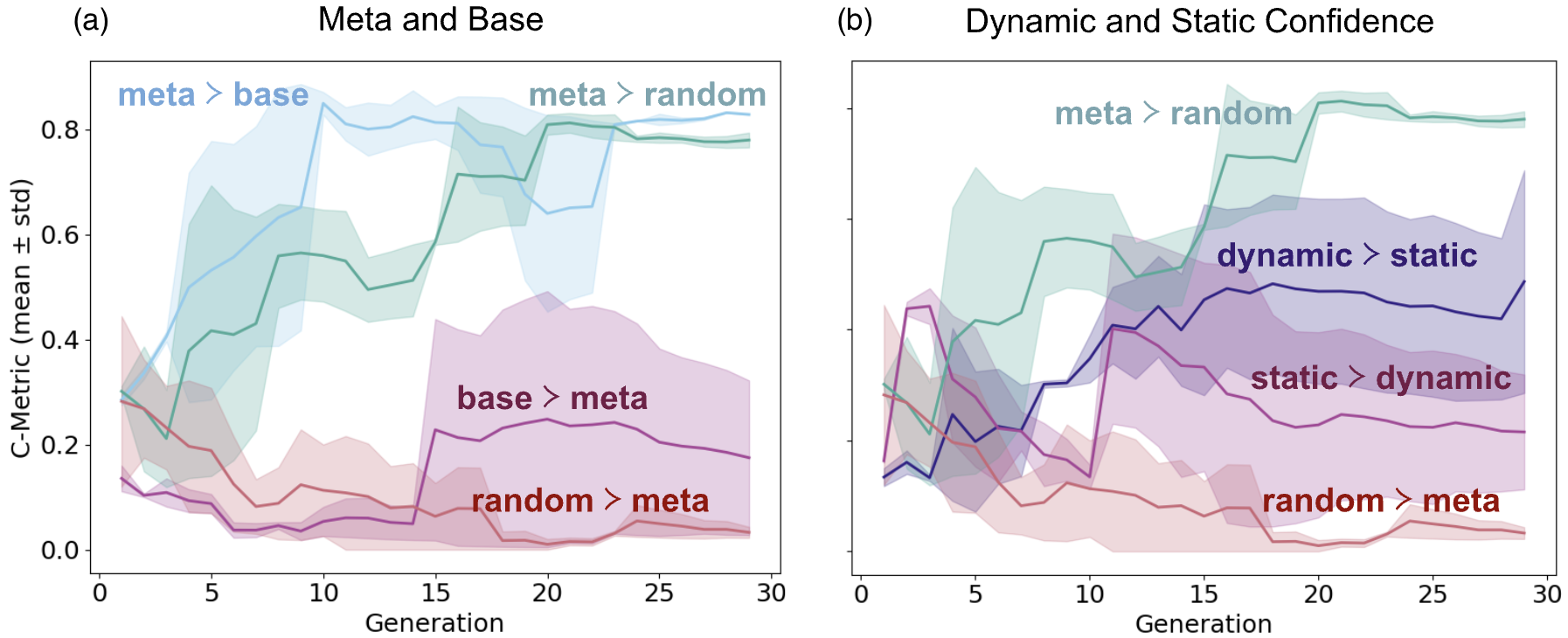}

    \caption{\textbf{Pareto-front dominance in the spin-crossover search.}
(a) C-metric values for the spin-crossover (SCO) experiment, reported as mean $\pm$ standard deviation over five runs. The meta-learning pipeline rapidly reaches C-metric values of 0.7--0.85 against the base-only pipeline and maintains this advantage across most generations. The reverse comparison remains much lower: the base-only pipeline rarely exceeds $C\sim0.25$ the meta-learning pipeline, and random selection remains near zero, indicating negligible dominance over meta-guided selection.
(b) C-metric comparison between meta-learning pipelines with static and adaptive uncertainty. By the final generation, approximately half of the points on the static-uncertainty Pareto front are dominated by at least one point on the adaptive-uncertainty front.}
    \label{fig:sco_cmetric}
\end{figure}

\begin{figure}[H]
    \centering
    \includegraphics[width=0.75\linewidth]{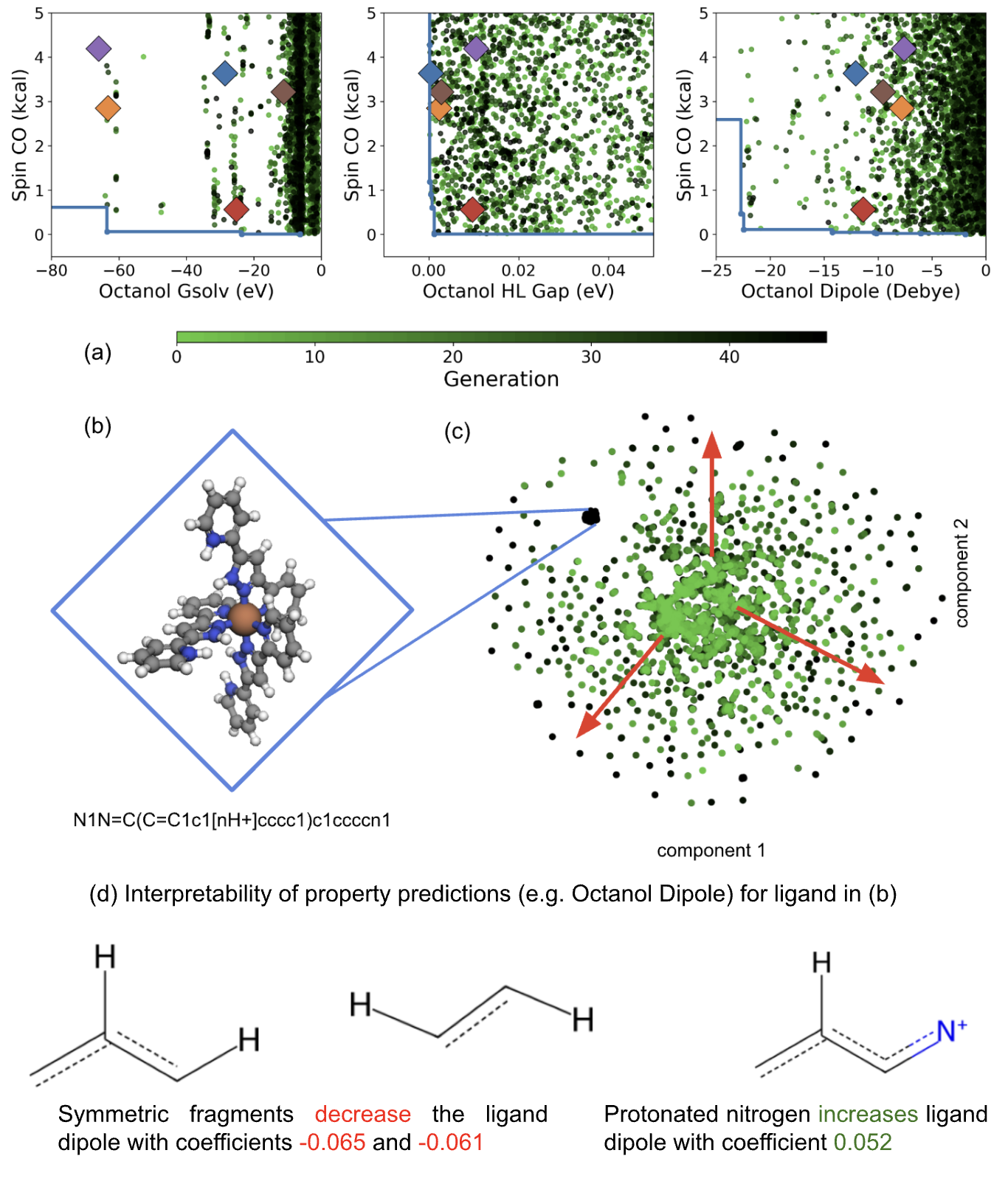}
    \caption{\textbf{Pareto-optimal complexes, chemical-space exploration, and model interpretability in the SCO experiments.}
    (a) Two-dimensional projections of the Pareto front from the largest search campaign, with all evaluated molecules colored by generation. The search targets complexes with $\Delta E_{\mathrm{SCO}} < 5$, as predicted by \texttt{Architector}, high solvation energy, high dipole moment, and low HOMO-LUMO gap, with all properties computed in octanol solution. Solvation energy and dipole moment are optimized through their sign-flipped objectives. The HOMO-LUMO gap forms a broad Pareto front, whereas the remaining objectives are dominated by a small number of extreme points. To select representative complexes from the full Pareto front, we rank candidates by favorable marginal percentiles across the objective distributions; the selected complexes are shown as colored diamonds.
    (b) Representative Pareto-optimal complex selected from the search.
(c) UMAP projection of the molecular fingerprints, with points colored by generation. The embedding shows both outward movement over successive generations, consistent with exploration, and dense regions revisited by the search, consistent with exploitation. One such region contains the blue-diamond complex highlighted in panels (a) and (b).
(d) Model interpretability. The largest-magnitude graphlet coefficients identify structural motifs that contribute most strongly to the predicted objectives.}
    \label{fig:pareto-frontier}
\end{figure}

The advantage of meta-guided selection is even clearer in Pareto dominance comparisons (Fig.~\ref{fig:sco_cmetric}a). Across five runs the meta pipeline produces a Pareto set that by the last iteration dominates on average \(78\%\) of the points in the base-only Pareto set, whereas only \(18\%\) of base Pareto points dominate the meta Pareto set. The meta strategy consistently dominates the base with $C(\mathrm{meta}, \mathrm{base})$ ranging from 70\% to 85\% throughout the search, while the reverse comparison $C(\mathrm{base}, \mathrm{meta})$ rarely exceeds 25\%. Although the SCO search space is too large to establish full convergence within the present evaluation budget, the dominance statistics show a persistent and substantial gain from meta-guidance. In a continued campaign, this advantage would likely translate into considerable wall-time savings by reducing the number of expensive evaluations needed to recover high-quality Pareto fronts.
In our largest SCO campaign, we started by randomly sampling a set of $M=500$ ligands from the base pool and evaluating the four target properties. After building the current Pareto front, each iteration fits Bayesian bootstrap ensembles of $B=100$ base models for all four objectives of the search, together with one additional unweighted model for each auxiliary task, using the updated set of evaluated molecules. This yields a total of $B\times N_o + N_a = 409$ ridge regressions per iteration that can be carried out completely in parallel. The pipeline then proposes $P=60000$ new molecules drawn from two sources: functionalized variants of ligands associated with current Pareto-front complexes and new ligands from the filtered CSD base pool. Using our ensemble of surrogates we produce a ranking using PoI. The $S=2000$ selected are subsequently clustered into $K=399$ representative molecules to evaluate, after which the next iteration begins. The presented workflow is highly modular and parallelizable. Apart from clustering, surrogate-based candidate ranking, and Pareto-front updates, the dominant steps are embarrassingly parallel: ensemble fitting, candidate generation, surrogate prediction, PoI evaluation, and high-fidelity property calculations can all achieve high degree of parallelization.Figure~\ref{fig:pareto-frontier}(a) shows the projected Pareto front obtained after 50 iterations of the largest campaign. 

The final Pareto set is large, containing $\sim 400$ complexes, yet remains smaller than the corresponding fronts produced by the base and random pipelines, which contain $\sim 500$ and $\sim 1100$ complexes, respectively. Further selection requires chemical judgment, since many Pareto-optimal complexes encode different trade-offs among the four objectives. As a simple chemically agnostic summary, we selected Pareto-front members with the highest average marginal percentile across the four properties.
Figure ~\ref{fig:pareto-frontier}c shows a two-dimenasional UMAP projection of the molecular fingerprints over the course of the search. Despite the complexity of visualizing in such a high dimensional space (from $\sim 10000$ fingerprint features to 2 dimensions), two trends are visible. Firstly there is an outward movement along generations, i.e. we are evaluating more and more diverse molecules (\emph{exploration}). Secondly, there are also cluster of points around which our pipeline focuses (\emph{exploitation}) on latter generations. One of these clusters corresponds to complexes similar to the blue colored diamond (plotted in Figure~\ref{fig:pareto-frontier}b).

\subsection{Models Performance}

\label{sec:models}

\begin{figure}[H]
    \centering
    \includegraphics[width=1.00\linewidth]{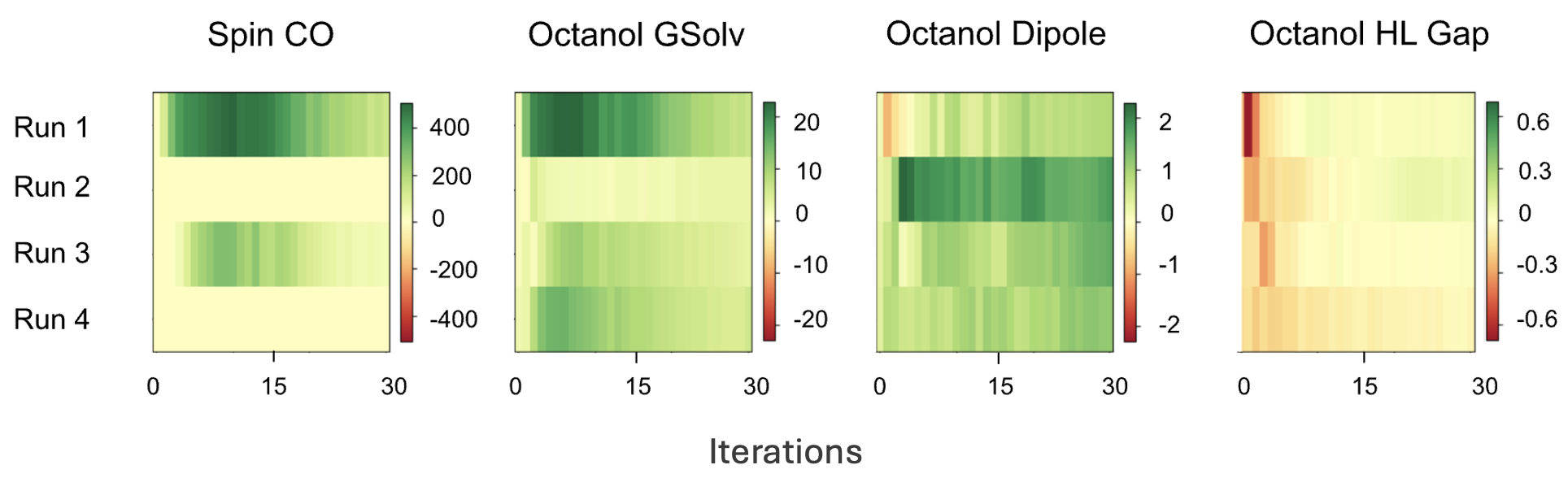}
    \caption{
    \textbf{Property prediction error on meta-selected SCO candidates over the course of search.}
Prediction error on generation $T+1$, evaluated over four independent runs after fitting the meta-learning and base models on all generations $\leq T$. Green regions indicate iterations where the meta-learning model has lower error, red regions indicate iterations where the base model has lower error. Across the 480 predictions shown, pooled over all objectives and iterations, the meta-learning model performs better in 91\% of cases, with the remaining cases showing comparable performance to the base model. Errors are evaluated on the molecules selected by the meta-learning pipeline. 
    }
    \label{fig:models-onestepahead}
\end{figure}

Figure~\ref{fig:models-onestepahead} compares the predictive accuracy of the linear models in a prospective setting. At generation $T+1$ the base and meta-learning models are fit using all molecules evaluated up to generation $T$. Across four independent runs, we report the relative RMSE difference between the base and meta-learning surrogates. The meta-learning models perform better in most cases, with the largest gains appearing early in the search, when only limited training data are available. This observation is consistent with the idea that meta-learning improves few-shot learning capabilities. The first three properties plotted show the clearest pattern, with meta-learning performing well when fitting with only $\sim$ 100 molecules and its benefit decreases as the iterations continue and training set increases. The largest observed gap reaches an RMSE difference of approximately (400) in the best run for the spin-crossover target, which is both the most expensive property to evaluate and the most difficult to predict. The HOMO-LUMO gap in octanol shows the weakest improvement, but meta-learning performance remains comparable to that of the base model. This shows empirically a similar result to what we expected theoretically: the meta algorithm is in the worst case very similar in terms of performance to just fitting its base linear models, as it might just resort to assign all the weights to the corresponding base linear model. The significant improvement in predictive performance comes with 2-3 times more computationally expensive fitting strategy compares to just a single regression, which still remains negligible relative to nonlinear deep learning models.\\ 

\begin{figure}[H]
    \centering
    \includegraphics[width=0.9\linewidth]{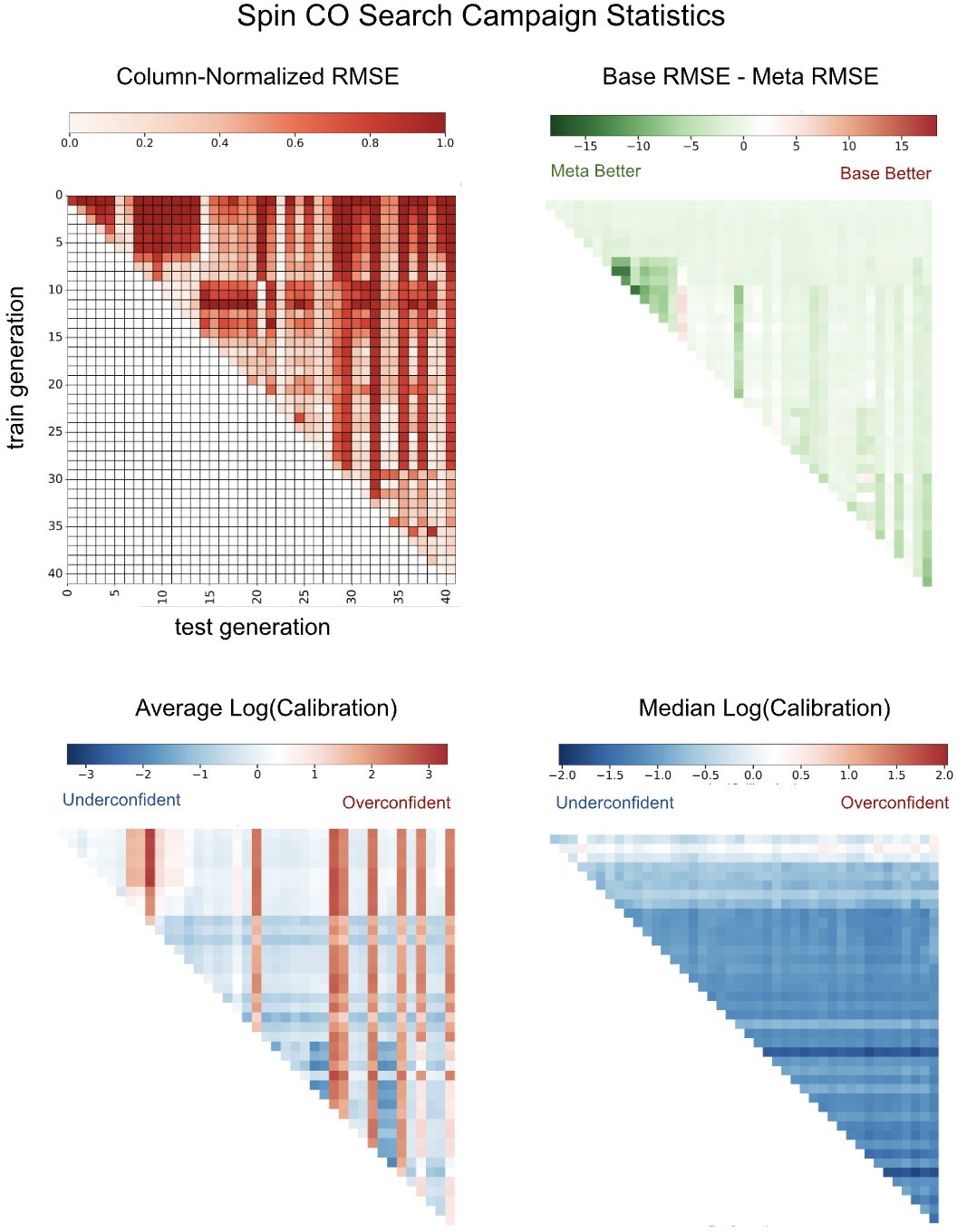}
    \caption{
    \textbf{Out-of-distribution robustness and uncertainty calibration for the spin-crossover target across training and testing generations}. The top-left panel shows column-normalized RMSE, the top-right panel reports the error difference between base and meta-learning models, and the bottom panels show calibration effects. Positive values indicate overconfidence, whereas negative values indicate underconfidence.
    }
    \label{fig:robustness-check}
\end{figure}

\noindent Figure~\ref{fig:robustness-check} reports an out-of-distribution robustness analysis for the spin-crossover target. Each entry in the triangular matrices corresponds to a train-test generation pair: the (y)-axis gives the last generation included in training, and the (x)-axis gives the generation used for testing. Moving downward expands the training set, whereas moving right evaluates the model farther into future generations. The top-left panel shows the column-normalized RMSE.
Normalizing by column highlights how performance changes as more generations are incorporated into training. Many columns show the expected transition from darker to lighter shades, indicating improved generalization as more data become available or effective learning along generations. However, some columns display persistently high errors or irregular patterns, revealing specific generations that are intrinsically harder to predict. The top-right panel reports the RMSE difference between the base and meta models. Green entries indicate lower error for the meta-learning model, and red entries indicate lower error for the base model. Across most training and testing combinations, meta-learning improves predictive accuracy. Finally, the bottom panels assess uncertainty calibration using \( \phi = \log\!\left((x - \bar{x})^2/\sigma^2\right)\), reported as both mean and median over samples from the same generation. The mean log-calibration is strongly influenced by the same hard generations that produce large prediction errors, creating localized overconfident regions. 
In contrast, the median log-calibration is more stable and remains predominantly negative across the matrix, indicating that the model tends to be underconfident for typical samples.

\subsection{Confidence Dynamic Update}
\label{sec:results-dynamic}

\begin{figure}[H]
    \centering
    \includegraphics[width=\linewidth]{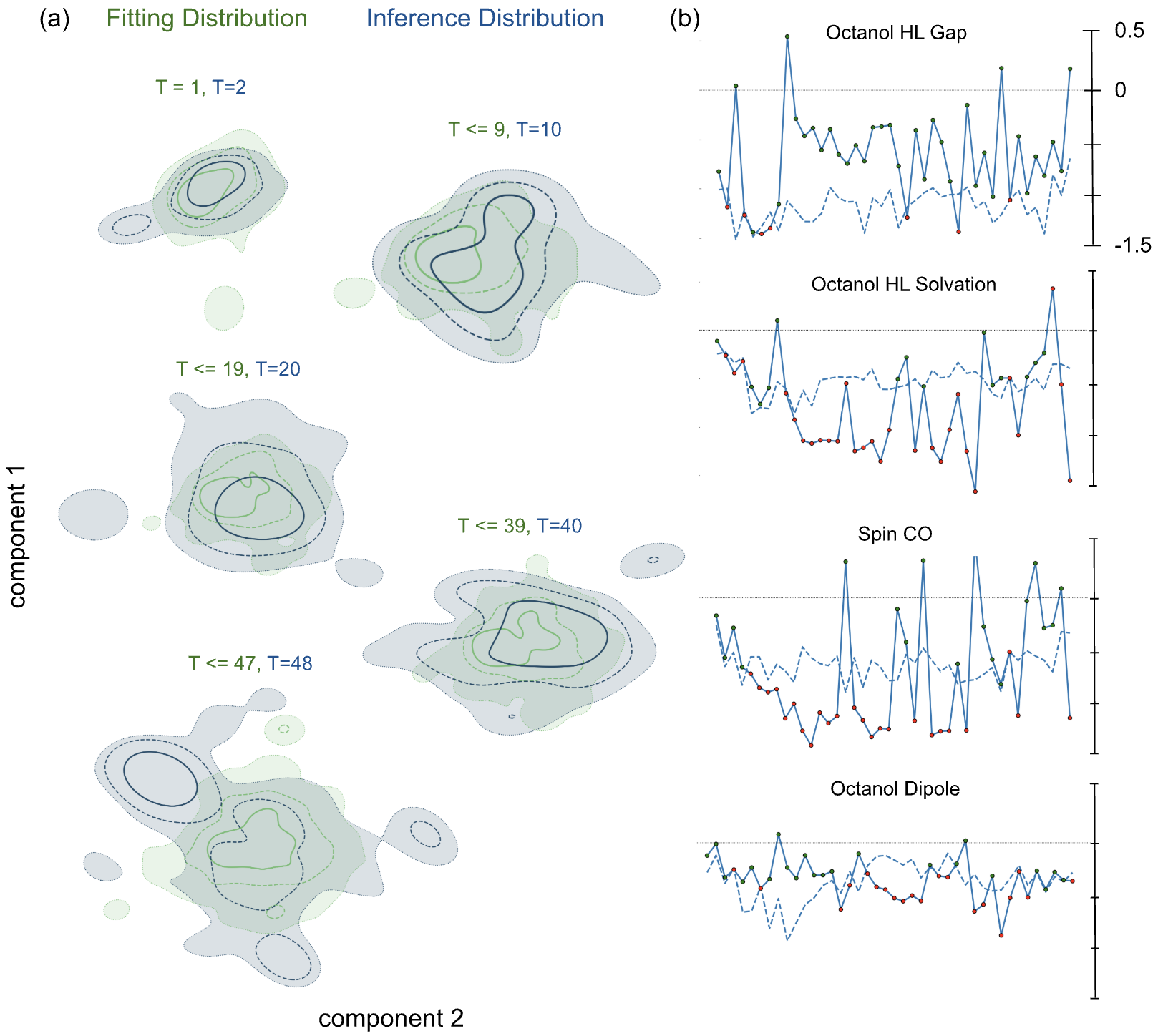}
    \caption{
    (a) Evolution of the evaluated molecule distribution in a two-dimensional PCA embedding of fingerprint space, comparing fitting (green) and inference (blue) distributions across iterations.
    (b) One-step-ahead calibration error for fixed and dynamically updated confidence parameters across objectives. Green points indicate iterations where dynamic confidence gives better median calibration; red points indicate iterations where fixed confidence is better.
    }
    \label{fig:calibration-plots}
\end{figure}

Figure~\ref{fig:calibration-plots}(a) illustrates the gap between fitting and inference distributions using a two-dimensional PCA projection of the fingerprint vectors. The points shown are not the full set of proposed candidates, but only the selected and evaluated ones. Nonetheless, we are able to capture the complexity of learning and calibrating during a search campaign. Beyond the continual expansion of the sampled chemical space, many inference distributions exhibit tail behavior or bimodality that is absent in the training data, making accurate prediction and balanced calibration challenging. In Figure~\ref{fig:calibration-plots}(b) we show one-step-ahead calibration for the static and dynamic confidence schemes, with the horizontal line corresponding to a perfect average calibration of $0$. Green points indicate generations in which the dynamic confidence update is closer to perfect average calibration than the static scheme; red points indicate the reverse. This comparison alone is not conclusive, since the two schemes alter the exploration-exploitation balance and select different molecules for evaluation. The resulting calibration values are therefore endogenous to the search trajectory: each confidence rule changes the candidate distribution on which it is later assessed.
For this reason, we focus on the empirical effect of the confidence update on the resulting Pareto fronts. Figure~\ref{fig:sco_cmetric}b compares the full pipeline with and without dynamic confidence adaptation. By the final iteration, 52\% of the points on the static-confidence Pareto front are dominated by at least one point on the dynamic-confidence front, whereas only 21\% of the dynamic-confidence front is dominated by the static-confidence front. This gap widens over the search, suggesting that the dynamic update recovers from a suboptimal initialization within a small number of iterations and yields better Pareto-front quality by the end of the campaign.

\subsection{Ablation Study}
\label{sec:ablation}

\begin{figure}[H]
        \centering
        \includegraphics[width=0.75\linewidth]{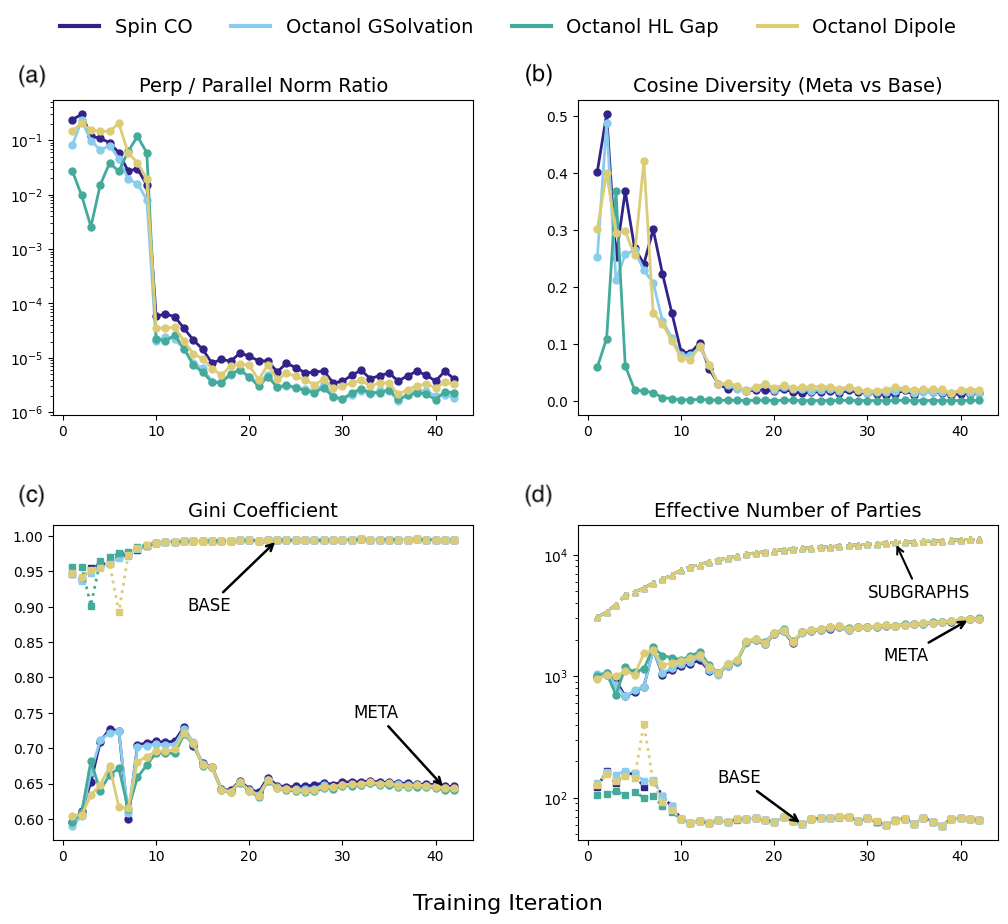}
        \caption{
        Meta-learning ablation across SCO objectives, showing parameter-vector geometry, coefficient inequality, and the effective number of contributing molecular subgraphs during training.
        }
        \label{fig:ablation}
\end{figure}

To better understand the mechanism underlying the performance gains from meta-learning, we perform an ablation study that tracks the learned coefficients and their geometry during training. A first clear effect of meta-learning appears in the Gini coefficient ~\cite{ceriani2012origins} (Fig.~\ref{fig:ablation}c) and the inverse Simpson index~\cite{simpson1949measurement}(Fig.~\ref{fig:ablation}d) of the learned coefficients. The base model has highly unequal coefficient magnitudes (average Gini $\approx 0.97$ across objectives), indicating that its predictions are dominated by a small number of molecular subgraphs. The meta model reduces this value to $\approx 0.67$, reflecting a broader distribution of weight across substructural features. The same trend appears in the inverse Simpson index, which measures the effective number of contributing subgraphs. Across objectives, the meta-learning model relies on roughly one order of magnitude more subgraphs than the base model. Moreover, the effective number of active subgraphs in the meta model grows along with dataset diversity, whereas the base model remains confined to a relatively small subset of predictors. These results suggest that meta-learning acts as an effective chemically-aware regularizer: instead of concentrating weight on a few predictors, it distributes importance across a broader set of chemically meaningful substructures, thereby mitigating overfitting in the low-data regime. The top panels provide a geometric interpretation of the relationship between meta and base parameter vectors. As training progresses and more data become available, the performance gap between meta and base models narrows, as expected in the large-data regime (Fig. ~\ref{fig:ablation}b). Decomposing the meta-learning parameter vector into components parallel and orthogonal to the base solution gives a complementary view of this convergence. The orthogonal-to-parallel norm ratio decreases rapidly during training, indicating that the orthogonal component becomes negligible as the model acquires more task-specific data (Fig. ~\ref{fig:ablation}a).

\clearpage

\section{Conclusions and Future Work}
\label{sec:conclusions}

We have presented a modular pipeline for multi-objective molecular discovery that combines interpretable linear meta-learning models, Bayesian uncertainty-driven active learning, and a dynamic confidence-tuning mechanism. Evaluated on QM9 and on a large-scale search for spin-crossover metal-organic complexes, the pipeline consistently outperforms both random search and non-meta baselines.

\noindent On QM9, both meta and base model-guided strategies converge nearly two orders of magnitude faster than random search, with the meta pipeline dominating the base Pareto front in 20-30\% in of cases before both converge to equivalent fronts. In spin-crossover complexes search, the advantage of meta-learning is markedly larger and more persistent with no sign of convergence within the computational budget. Several factors likely contribute to this performance difference. First, the underlying property landscapes are intrinsically different: SCO targets exhibit a more complex and irregular structure-property relationship than the QM9 objectives. Second, the sampling regimes are distinct. QM9 candidates are drawn approximately i.i.d. from a small, homogeneous molecular space, whereas SCO candidates are generated partly by functionalizing previously discovered Pareto-optimal complexes. This mutation-driven, non-i.i.d. sampling process is precisely the regime in which meta-learning regularization is most valuable. Third, the SCO candidate space is orders of magnitude larger than QM9, making efficient generalization much more important.

\noindent The dynamic confidence-tuning algorithm yields empirically better Pareto fronts. By the final generation, more than 50\% of the Pareto points selected under static confidence are dominated by the dynamic strategy. Its mechanism, though, is difficult to characterize through calibration alone. The calibration metric alone does not cleanly distinguish the two strategies, because the two algorithms implicitly evaluate molecules of different difficulty by steering the exploration-exploitation trade-off in different directions along the iterations. Each confidence value encodes a distinct point on the curve between full exploration and full exploitation, and our dynamic update oscillates between these regimes in a way that a fixed value cannot. The difficulty of assessing uncertainty calibration during an active search campaign is not unique to our method: it is a fundamental challenge of adaptive experimental design. Developing principled calibration metrics and effective uncertainty quantification for this setting is, in our view, an important open problem. 

\noindent The ablation study yields interesting insights into advances in linear meta-learning. While it was predictable that meta-learning would act as a regularizer, the mechanism through which it does so is different from what one might have expected. Rather than selecting a sparse subspace of generally predictive features, meta-learning regularizes by broadening the set of features actively used during prediction. This redistribution of weight across chemically meaningful features appears to reduce dependence on a small number of high-magnitude coefficients and improves generalization in the low-data regime.

\noindent Our pipeline has a potential limitation worth mentiontioning: for large search campaigns it produces big Pareto sets. It is direct consequence of carrying out search campaigns in high dimensions and an instance of the so-called curse of dimensionality ~\cite{bellman2015adaptive}, a widely known phenomenon without an engineering solution free of assumptions. Yet this behavior can also be viewed as a design choice for human-in-the-loop discovery. Rather than imposing an early scalarization that compresses the objectives into a single score, the pipeline preserves a diverse set of non-dominated trade-offs and leaves the final selection to chemical judgment. 

\noindent A practical strength of the pipeline is its modularity. The meta-learning surrogate, the Bayesian bootstrapping uncertainty quantification, and the dynamic confidence update are loosely coupled components that can each be replaced or improved independently.

\noindent Three directions stand out as natural extensions.  First, the pipeline currently evaluates all objectives for every selected candidate, regardless of their relative computational cost. Decoupling the acquisition decisions per objective, so that expensive properties are evaluated only when the surrogate uncertainty justifies it, would substantially reduce wall-time in settings where objectives differ in cost by orders of magnitude; cost-aware multi-objective acquisition strategies such as PESMO~\cite{HernandezLobato2016PESMO} offer a starting point, though extending them beyond Gaussian process surrogates remains an open challenge. Second, applying the pipeline to additional domains would test the generality of the approach and potentially reveal new failure modes and opportunities for improvement. Finally, it would be of interest to expand the set of auxiliary tasks, for instance evaluating cheap properties from widely available packages like RDKit or considering other side-products of computational chemistry.

\section{Code and Data availability}
The code is available on Github together with instructions on how to reproduce the experiments ~\cite{RepoMinerva}. The models and pipelines will be integrated as part of the \href{https://github.com/tonioeltopoquegira/minervachem}{$\texttt{minervachem}$} package. The data produced by the experiments are uploaded on Zenodo ~\cite{ZenodoRepo}.

\section{Acknowledgment}

This work is supported by Los Alamos National Laboratory LDRD and NSF awards No. ~IIS~2345860 and ~IIS~2435905.

\printbibliography
\newpage

\section{Supplementary Information}

Results reported in Fig. ~\ref{fig:qm9_cmetric}, Fig. ~\ref{fig:orders-1D} are averages and standard deviations for 10 QM9 experiment runs. Fig. ~\ref{fig:sco_cmetric} and Fig. ~\ref{fig:models-onestepahead} are ensembles of 4 runs for the SCO experiment experiment. 

\subsection{Architector Experiment Specifications}

\paragraph{Pool Ligands Construction} Starting from an initial collection of approximately \( \sim 106{,}000 \) ligands. We exclude sandwich and edge-bridging ligands and retain only ligands with \(\{1,2,3,6\}\) denticities. In addition, we keep only ligands that would yield total complex charge bigger than $-3$. After filtering, the resulting base ligand pool contains approximately \( \sim 50{,}000 \) ligands. 

\paragraph{Experiment Parameters} For the QM9 benchmark experiment, the pipeline was initialized with
$M=40$ molecules, proposing $P=1,000$ candidates per iteration, retaining $S=10$ to look-up, with $B=100$ bootstrap samples and run for $100$ iterations. No clustering is applied. For the main ligand search campaign, the pipeline was initialized with $M=500$ ligands, proposing $P=60000$ candidates per iteration, with the top $S=2000$ clustered into $K=399$ final molecules for evaluation, $B=200$ bootstrap samples, and run for $50$ iterations.

\paragraph{Hardware specification} On average, generating a conformer and evaluating the spin-state energies with Architector requires approximately 11 minutes on a single AMD CPU core. Various AMD CPU models were used but most frequently the AMD EPYC 9654. Maximum wall-time per Architector evaluation is set to 20 minutes. After that the code is returned and the construction of ligand-center complex is considered as failed.

\subsection{Robustness Check for other properties}

\begin{figure}[H]
    \centering
    \includegraphics[width=0.67\linewidth]{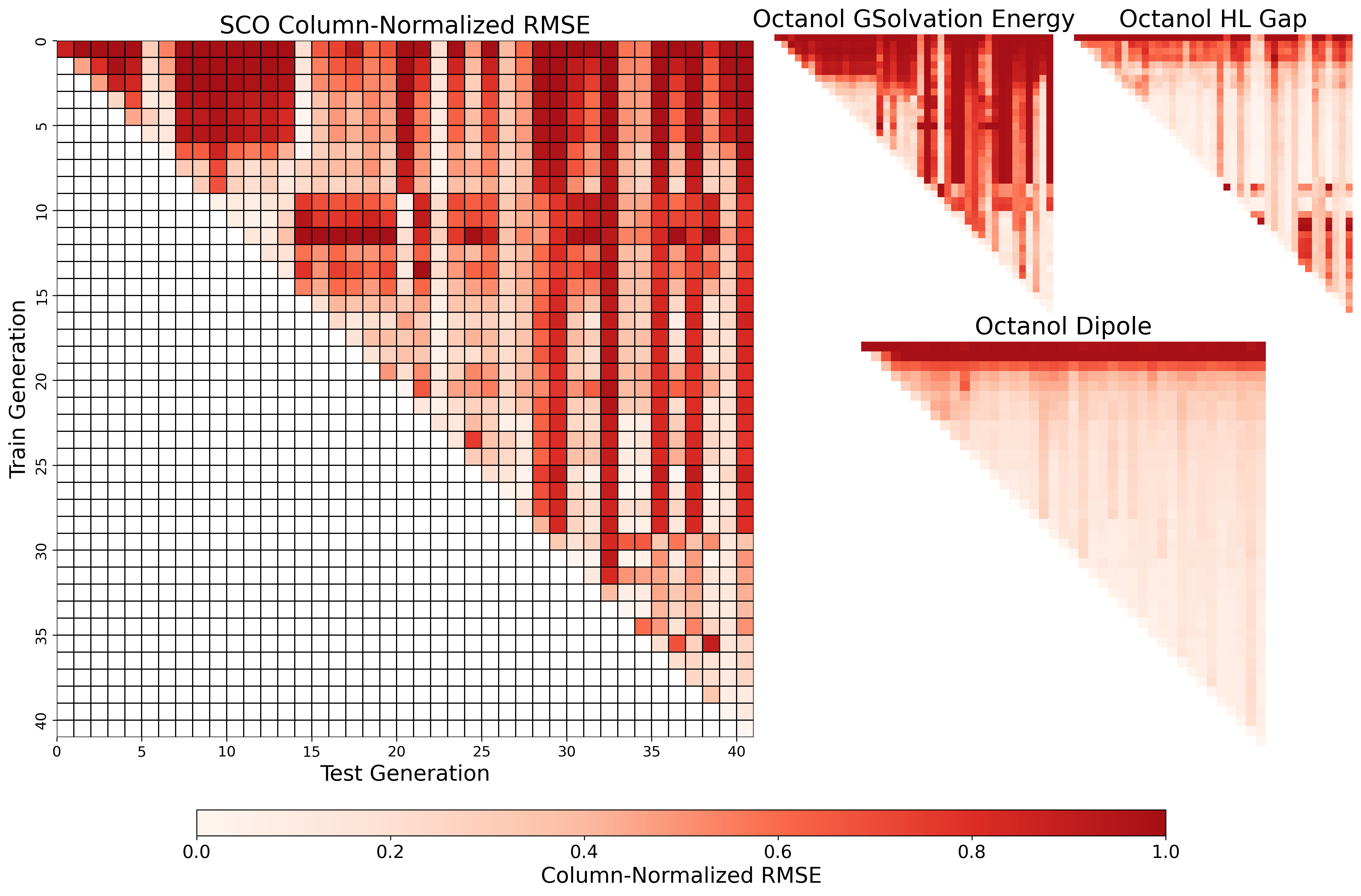}
    \caption{Column-normalized RMSE for each property.}
    \label{fig:placeholder}
\end{figure}

\begin{figure}[H]
    \centering
    \includegraphics[width=0.67\linewidth]{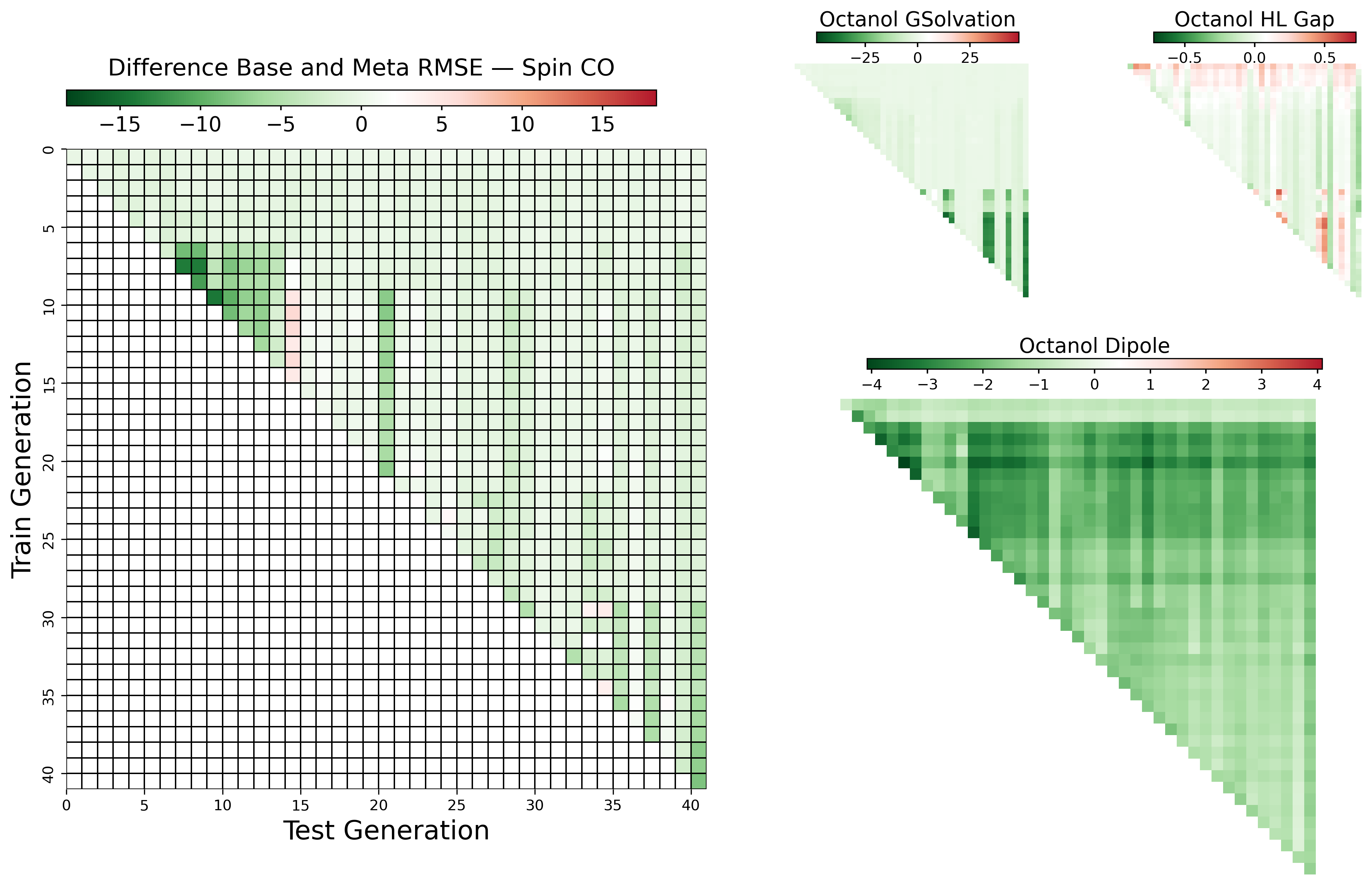}
    \caption{RMSE error difference for each property.}
    \label{fig:placeholder}
\end{figure}

\begin{figure}[H]
    \centering
    \includegraphics[width=0.67\linewidth]{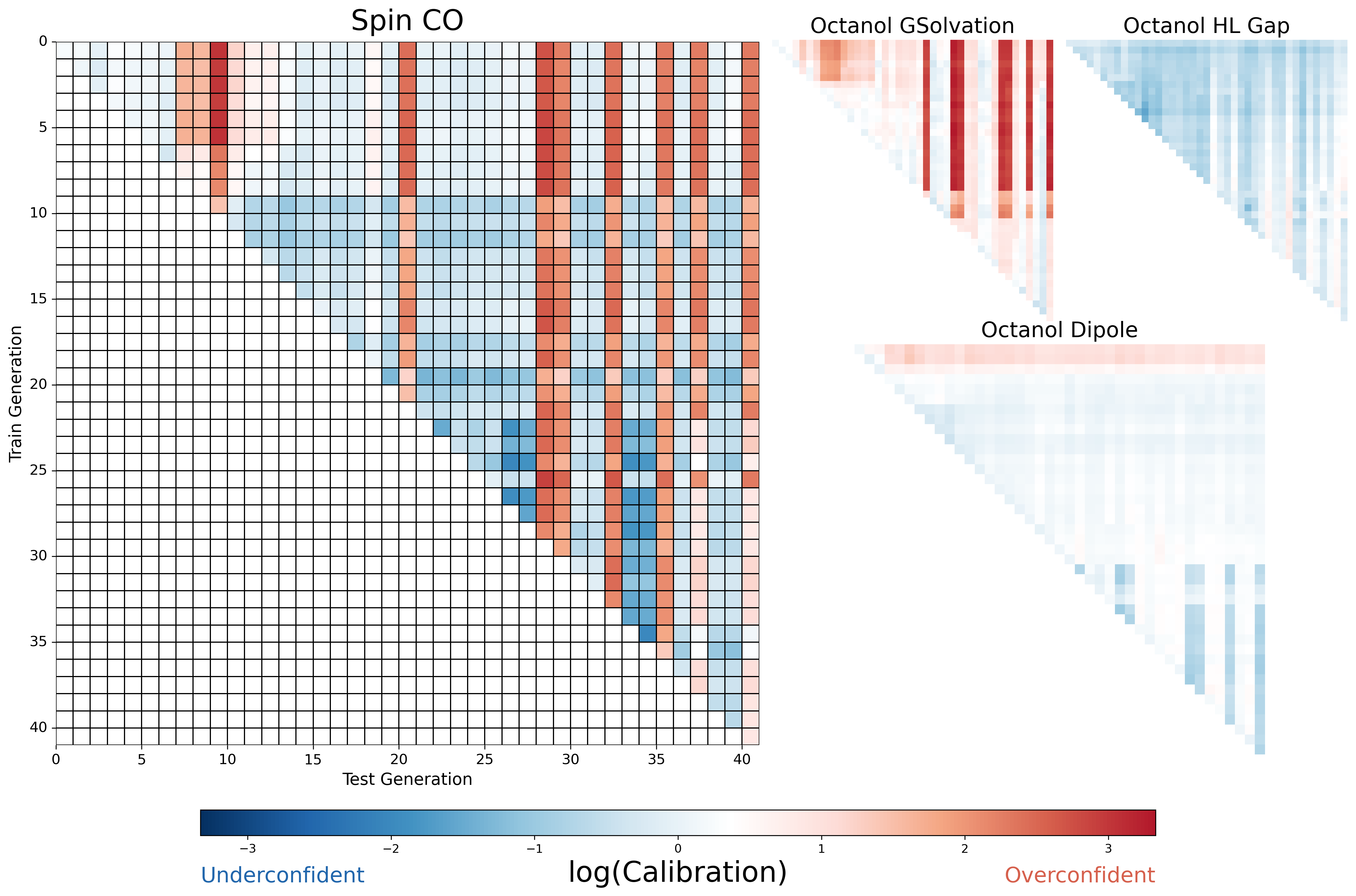}
        \caption{Mean Calibration for the linear meta model for each property}
    \label{fig:placeholder}
\end{figure}

\begin{figure}[H]
    \centering
    \includegraphics[width=0.67\linewidth]{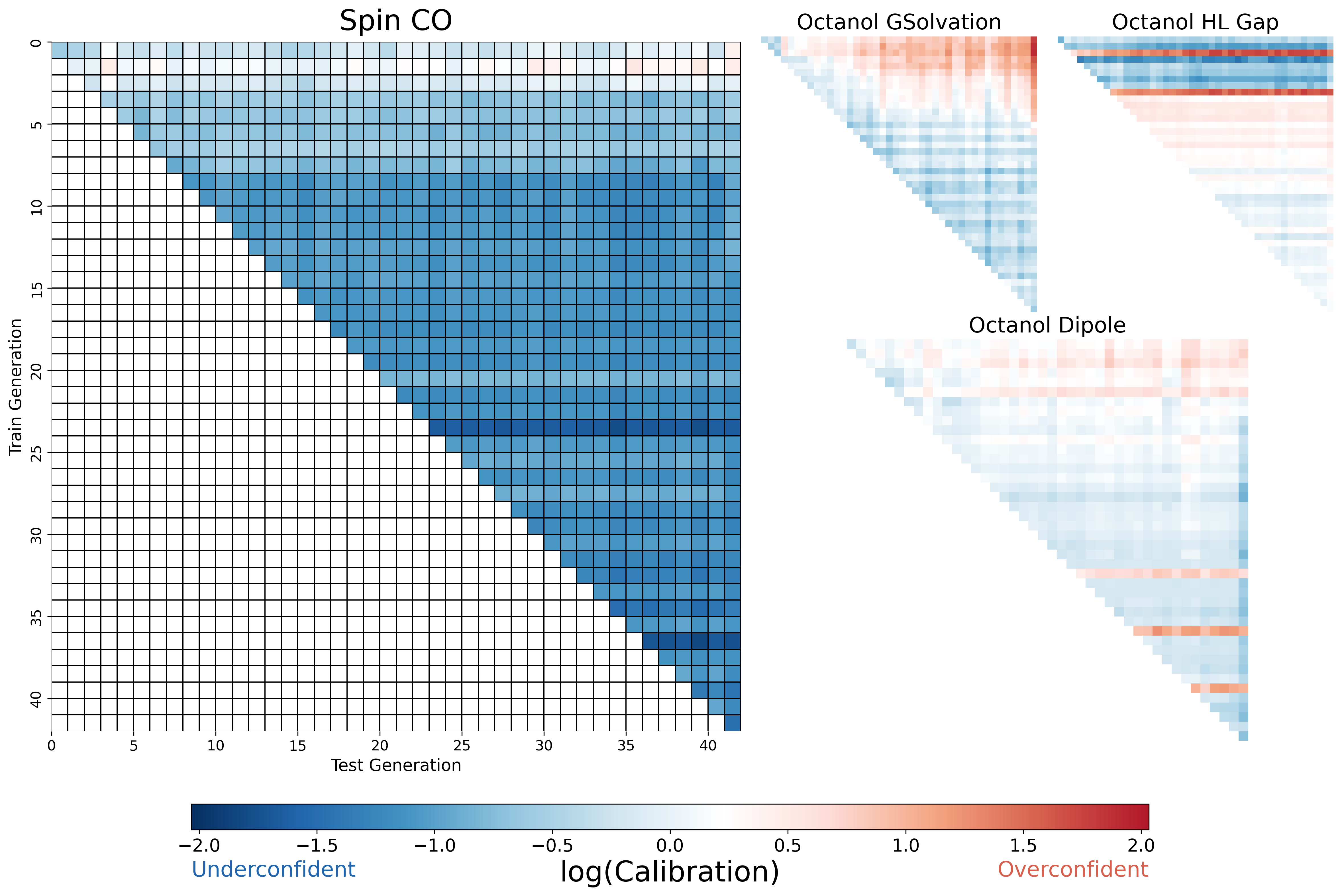}
        \caption{Median Calibration for linear meta model for each property}
    \label{fig:placeholder}
\end{figure}

\end{document}